\documentclass[prd,twocolumn,10pt,aps,longbibliography]{revtex4-2}

\usepackage{graphicx}% Include figure files
\usepackage{dcolumn}% Align table columns on decimal point
\usepackage{bm}% bold math
\usepackage[T2A]{fontenc}
\usepackage{comment}
\usepackage{braket}
\usepackage{amsmath}
\usepackage{amssymb}
\usepackage{dsfont}
\usepackage{upgreek}
\usepackage{natbib}
\newcommand{\bfp}{{\bf p}}
\newcommand{\bfx}{{\bf x}}
\newcommand{\bfq}{{\bf q}}

\newcommand{\bfqq}{{\bf q'}}

\begin{document}

%\preprint{APS/123-QED}

\title{Quantum field-theoretic framework for neutrino decoherence from scattering in a medium}% Force line breaks with \\
%\thanks{A footnote to the article title}%

\author{Konstantin Stankevich}
% \altaffiliation[Also at ]{Physics Department, XYZ University.}%Lines break automatically or can be forced with \\
%\author{Second Author}%
\email{kl.stankevich@physics.msu.ru}
\affiliation{%
	Department of Theoretical Physics, Faculty of Physics,
	Lomonosov Moscow State University, 119992 Moscow, Russia
	% This line break forced with \textbackslash\textbackslash
}

%\collaboration{MUSO Collaboration}%\noaffiliation

\author{Alexander Studenikin}
\email{studenik@srd.sinp.msu.ru}
% \homepage{http://www.Second.institution.edu/~Charlie.Author}
\affiliation{%
	Department of Theoretical Physics and Department of Physics of Particles and Extreme States of Matter, Faculty of Physics, Lomonosov Moscow State University, 119992 Moscow, Russia
	% This line break forced with \textbackslash\textbackslash
}

\author{Maksim Vyalkov}
% \altaffiliation[Also at ]{Physics Department, XYZ University.}%Lines break automatically or can be forced with \\
%\author{Second Author}%
\email{vyalkovmsu@yandex.ru}
\affiliation{%
	Department of Theoretical Physics, Faculty of Physics,
	Lomonosov Moscow State University, 119992 Moscow, Russia}
  \affiliation{  Branch of Lomonosov Moscow State University in Sarov, 607328 Russia} 
   \affiliation{ National Centre for Physics and Mathematics, Sarov 607182, Russia}
    
	% This line break forced with \textbackslash\textbackslash

\date{\today}% It is always \today, today,
%  but any date may be explicitly specified

\begin{abstract}

We develop a theoretical framework for quantum field description of neutrino evolution in a medium, with a focus on quantum decoherence induced by neutrino scattering on fermions. By deriving a generalized Lindblad master equation that accounts for neutrino momentum-changing transitions, we go beyond the standard treatment that assumes fixed neutrino momentum. Our model explicitly connects decoherence parameters to  scattering cross sections, offering a bridge between theoretical predictions and experimental constraints on neutrino quantum decoherence. We apply the developed formalism to several scenarios: 1) neutrino scattering on electrons, 2) neutrino scattering on protons and neutrons through non-standard interactions (NSI), 3) neutrino scattering on dark matter fermions. The obtained results demonstrate that decoherence effects can significantly alter neutrino oscillation patterns and provide a new probe for physics beyond the Standard Model.

\end{abstract}

\pacs{Valid PACS appear here}% PACS, the Physics and Astronomy
% Classification Scheme.
%\keywords{Suggested keywords}%Use showkeys class option if keyword
%display desired
\maketitle

%\tableofcontents

\section{\label{sec:level1} Introduction}

Neutrinos have always provided a window to new physics for about a century \cite{Giunti:2014ixa}. The study of neutrino decoherence, an effect that occurs when neutrinos interact with an external environment, has opened up a new and very promising field in neutrino physics. This phenomenon can be described by the Lindblad master equation \cite{Lindblad:1975ef,Gorini:1975nb}, which takes the form:

\begin{equation}\label{Lindblad_eq}
    \dfrac{\partial \rho_\nu(t)}{\partial t} = - i \left[  H(t),\rho_\nu(t) \right] + \sum_i \Gamma_i\left(L_i \rho_\nu L_i^{\dagger}-\frac{1}{2}\left\{L_i^{\dagger} L_i, \rho_\nu\right\}\right),
\end{equation}
where $\rho_\nu(t)$ is the neutrino density matrix, $H(t)$ is the neutrino Hamiltonian, and $L_i$ and $\Gamma_i$ are the Lindblad operators and decoherence parameters, respectively, which characterize the dissipative effects. It is important to note that the Lindblad equation specifies the general structure of the evolution but does not predetermine the explicit form of the dissipative operators or the values of the parameters. Additionally, the Lindblad formalism is typically applicable for describing decoherence resulting from transitions between different quantum states with the same momentum, a condition that may not always hold for propagating neutrinos.

Searches for evidence of neutrino quantum decoherence are conducted using neutrino fluxes from various sources. The quantum decoherence effect has been studied for reactor and accelerator neutrinos neutrinos  \cite{BalieiroGomes:2016ykp, Oliveira:2016asf, Coelho:2017byq, deGouvea:2021uvg, deGouvea:2020hfl, DeRomeri:2023dht,Bera:2023ksx,Bera:2024hhr,ESSnuSB:2024yji}, for the solar neutrinos \cite{deHolanda:2019tuf,Fogli:2007tx}, and for atmospheric neutrinos  \cite{KM3NeT:2024jji,IceCube:2025zim,Coloma:2018idr}. It has been suggested that neutrino quantum decoherence might offer an explanation for the gallium neutrino anomaly \cite{Farzan:2023fqa}.

Recent studies \cite{Ternes:2025mys,dosSantos:2023skk} have indicated that neutrino quantum decoherence could also influence neutrino fluxes from supernova bursts, suggesting that supernova neutrinos provide another viable context for this search. In particular, the impact of quantum decoherence on collective neutrino oscillations has been explored in \cite{Purtova:2024lvy,Purtova:2023hcy, Stankevich:2023qdi, Stankevich:2020sja}.

Nevertheless, despite a considerable number of studies in this field,  the fundamental mechanisms leading to neutrino quantum decoherence remain not fully understood. In \cite{PhysRevD.50.4762, Burgess:1996mz,PhysRevD.71.013003, Lichkunov:2025fgo}, decoherence induced by classical fluctuations of the external environment was considered. We also note \cite{PhysRevD.104.043018} that spacetime fluctuations could give rise to neutrino quantum decoherence. The authors of \cite{Nieves:2020jjg,Nieves:2019izk} presented a study on decoherence mechanisms resulting from neutrino scattering. Previously, we developed a formalism to describe neutrino evolution accounting for quantum decoherence due to neutrino decay \cite{Stan2025,PhysRevD.101.056004} based on the framework of quantum electrodynamics of open systems \cite{Breuer:2007juk,Brasil:2012trs}.

We also note that many authors \cite{bowen2024opendynamicsinteractingquantum, Fogedby:2026zjj,Koksma_2010,burrage2025fockstateprobabilitychanges,Breuer:2007juk,Nandi:2025qyj,Boyanovsky:2015xoa,Burrage:2025xac,Huggett:2025pxz,Kading:2025cwg} are actively formulating and developing a new branch of quantum field theory, which can be called  "open quantum field theory" or "physics of open quantum fields."

In this paper, we further develop our quantum field-theoretic framework  \cite{Stan2025,PhysRevD.101.056004} to describe neutrino evolution, accounting for the neutrino interaction with an external environment. Within this approach we consider a new mechanism of neutrino quantum decoherence that arises from neutrino scattering on fermions. We derive the general master equation for neutrino evolution and consider several specific cases: 1) neutrino scattering on electrons, 2) neutrino scattering on protons and neutrons through non-standard interactions (NSI), 3) neutrino scattering on dark matter fermions. We also show that the obtained neutrino master equation describes such effects as neutrino quantum Zeno effect and Brownian neutrino motion that includes neutrino transitions.

\section{EVOLUTION MASTER EQUATION}

\subsection{General case}

To obtain the evolution equation for the neutrino density matrix, we follow the formalism developed in \cite{Stan2025}. Consider a Dirac neutrino described by a density matrix:

\begin{equation}
\rho_\nu (t) = \ket{\Phi(t)} \bra{\Phi(t)},
\end{equation}
where the state $\ket{\Phi(t)} = \otimes_{\mathbf{p}} \ket{\Phi_{\mathbf{p}}}$ is a tensor product over states with different momenta. Our aim is to derive an evolution equation for the neutrino that accounts for transitions between its stationary states, accompanied by the scattering on fermion particles. 

We focus on the density matrix elements:
\begin{equation}
\rho_\bfp^{ij}(t) = \bra{i \bfp} \rho_\nu (t) \ket{j \bfp}.
\end{equation}
where  $\ket{i \bfp} = \sqrt{2 E_{\bfp i}} a_{\bfp i}^{\dagger} \ket{0}$ is a neutrino state  $i$ (that takes into account the neutrino helicity) with momentum $\bfp$. We assume there is no coherence between states of different momentum, meaning the off-diagonal momentum elements vanish:
\begin{equation}\label{diag_p}
\bra{i \bfp} \rho_\nu(t) \ket{j \bfq} = (2 \pi)^3 \delta^{(3)}(\bfp-\bfq) \rho_\bfp^{ij}(t).
\end{equation}

We consider the neutrino propagation in a bath of fermions in an equilibrium state. The evolution of the full system's density matrix (neutrinos and fermionic environment) $\varrho \left( t\right)$ in the interaction representation is given by the evolution operator $U(t_0, t)$:
\begin{equation}\label{rho}
\varrho \left( t\right) = U(t_0, t) \: \varrho (t_0) \: U^{\dagger}(t_0, t).
\end{equation}
We assume that these systems are initially weakly coupled. Therefore the density matrix of the full system at the initial time is expressed as
\begin{equation}
\varrho(t_0) = \rho_\nu (t_0) \otimes \rho_E (t_0),
\end{equation}
where $\rho_E (t_0)$ is the environmental density matrix.
The evolution operator is defined as follows
\begin{equation}
U(t_0, t) = T\: exp \left( -i \int^t_{t_0} H_I(t') dt' \right).
\end{equation}

The interaction Hamiltonian between the neutrino and the environmental fermions (e.g., electrons) is given in a general form:
\begin{equation} \label{Hamiltonian}
H_I(t) = \int d^3 \bfx j_{\mu}(x) J^{\mu}(x),
\end{equation}
where $J_{\mu}(x)$ is the medium current (e.g., $J_{\mu} (x) = \bar{\psi}_E(x) \Gamma^{\prime}_{\mu} \psi_E(x)$), and $j_\mu(x) = \bar{\nu}(x) \Gamma_\mu \nu (x)$ is the neutrino current ($\Gamma^{\prime}_\mu$ and $\Gamma_\mu$ are corresponding interaction vertices). The neutrino field operator are expressed as:
\begin{equation}
    \nu (x) = \sum_i \int \dfrac{d^3 \bfp }{(2 \pi)^3} \dfrac{1}{\sqrt{2 E_{\bfp i}}} a_{{\bf p} i} u_i(\bfp) e^{-i p x}
\end{equation}
Here, the negative-frequency solutions with the antineutrino creation operator were omitted, e.g. we assume that the neutrino interactions do not induce transitions between neutrino and antineutrino. The 
environmental field operator is also expanded in the standard plane-wave 
basis:
\begin{equation}
      \psi (x) = \int \dfrac{d^3 \bfp }{(2 \pi)^3} \dfrac{1}{\sqrt{2 E_{\bfp}}} a_{{\bf p}} v(\bfp) e^{-i p x}.
\end{equation}

Following the derivation similar to \cite{Stan2025}, we obtain the equation for the density matrix:

\begin{widetext}
\begin{equation}\label{equation_for_rho}
\begin{aligned}
    \dfrac{\partial \rho_\nu (t)}{\partial t} =  - i \int d^3 \bfx \left[j_\alpha(x) \braket{J^\alpha(x)}, \rho_\nu(t_0) \right]
    + \int d^3 \bfx_1 \int_{t_0}^{t_1} d^4 x_2  \braket{J^{(2) \alpha} J^{(1)\beta}} (j_{\alpha}^{(2)} \rho_\nu(t) j_\beta^{(1)} - j^{(1)}_\alpha j_\beta^{(2)}\rho_\nu(t))  +\\
    +   \int d^3 \bfx_1 \int_{t_0}^{t_1} d^4 x_2  \braket{J^{(1)\alpha} J^{(2)\beta}} (j^{(1)}_{\alpha} \rho_\nu(t) j_\beta^{(2)} -  \rho_\nu(t)j_\alpha^{(2)}j_\beta^{(1)}) ,
\end{aligned}
\end{equation}
\end{widetext}
where we use the shorthand notations $j_\alpha^{(1)} = j_\alpha(x_1)$, $j_\alpha^{(2)} = j_\alpha(x_2)$, $J_\alpha^{(1)} = J_\alpha(x_1)$, $J_\alpha^{(2)} = J_\alpha(x_2)$, and angle brackets denote the average.
To obtain the master equation for neutrino evolution, we must calculate the correlation function $\braket{J^\alpha(x_2) J^\beta(x_1)}$. We average according to the equilibrium state of the external environment:

\begin{equation}
    \langle \hat{O} \rangle = tr_E\left(\hat{O} \frac{1}{Z} \exp [-H_E/T]\right),
\end{equation}
where $H_E$ is the free Hamiltonian of the medium,  $T$ and $Z$ are environmental temperature and partition function correspondingly. 

Further we write equation (\ref{equation_for_rho}) in terms of neutrino states

\begin{widetext}
\begin{equation}\label{equation_1}
\begin{aligned}
    \dfrac{\partial \rho^{ij}_\bfp(t)}{\partial t} = 
     \sum_{n,n'} \int d^3 q \int d^3 q'  \int d^3 \bfx_1 \int_{t_0}^{t_1} d^4 x_2  
\langle J_{(2)}^{\alpha}J_{(1)}^{\beta}\rangle \bra{i \bfp}j^{(2)}_\alpha \ket{n \bfq} \bra{n \bfq} \rho_\nu(t) \ket{n' \bfqq} \bra{n' \bfqq} j^{(1)}_\beta \ket{j \bfp} - 
    \\
    - \sum_{n,n'} \int d^3 q \int d^3 q' \int d^3 \bfx_1 \int_{t_0}^{t_1} d^4 x_2 \langle J_{(2)}^{\alpha}J_{(1)}^{\beta}\rangle \bra{i \bfp} j_\alpha^{(1)} \ket{n \bfq} \bra{n \bfq} j_\beta^{(2)} \ket{n' \bfqq} \bra{n' \bfqq} \rho_\nu(t) \ket{j \bfp} +\\
   +  \sum_{n,n'} \int d^3 q \int d^3 q'  \int d^3 \bfx_1 \int_{t_0}^{t_1} d^4 x_2  
\langle J_{(2)}^{\alpha}J_{(1)}^{\beta}\rangle \bra{i \bfp}j^{(1)}_\alpha \ket{n \bfq} \bra{n \bfq} \rho_\nu(t) \ket{n' \bfqq} \bra{n' \bfqq} j^{(2)}_\beta \ket{j \bfp} - \\
- \sum_{n,n'} \int d^3 q \int d^3 q' \int d^3 \bfx_1 \int_{t_0}^{t_1} d^4 x_2 \langle J_{(2)}^{\alpha}J_{(1)}^{\beta}\rangle \bra{i \bfp} \rho_\nu(t) \ket{n \bfq} \bra{n \bfq} j_\alpha^{(2)}\ket{n' \bfqq} \bra{n' \bfqq} j_\beta^{(1)}  \ket{j \bfp},
\end{aligned}
\end{equation}
\end{widetext}
where we define 
$\int d^3 q  = \int \dfrac{d^3 \bfq}{2 (2\pi)^3 E_{\bfq n}}$,
$\int d^3 q'  = \int \dfrac{d^3 \bfqq}{2 (2\pi)^3 E_{\bfqq n'}}$. The matrix elements of the neutrino current are expressed in the following form

\begin{equation}\label{matrix_element}
    \bra{i \bfp}j_\alpha(x) \ket{j \bfp} = \bar{u}_{i}(\bfp) \Gamma_\alpha u_j(\bfp) e^{i (p-q) x }.
\end{equation}

To obtain a Lindblad-type equation, we first apply the rotating wave approximation (RWA). The rotating wave approximation in the quantum optical regime (also known as the secular approximation) involves retaining only the secular components in the double sums over the system frequencies and averaging out rapidly oscillating terms. 

The matrix element (\ref{matrix_element}) is proportional to the exponent $e^{i(E_p - E_q) t}$. The RWA is applicable when  $\Delta\omega \gg L$, where $L$ is the neutrino mean free path in the environment, and $\Delta\omega = \left|E_f - E_i\right|$ is the energy difference between neutrino states in the scattering process. For neutrinos with energies of the order of $1$ MeV (e.g. $\Delta\omega< 1$\:MeV) and external environment composed of electrons (and noting that $L \approx (n_e\sigma)^{-1}$, where $\sigma$ is the total cross section of neutrino scattering on electrons and $n_e$ is the electron density) we obtain the  electron density for RWA applicability should be $n_e < 10^{55}\text{cm}^{-3}$, which is many orders of magnitude greater than the electron density in a supernova. Therefore, the RWA is applicable to a wide range of astrophysical objects, including supernovae, stars, and accretion disks.

After applying the rotating-wave approximation, Equation (\ref{equation_1}) takes the form:

\begin{widetext}
\begin{equation}\label{equation_2}
\begin{aligned}
    \dfrac{\partial \rho^{ij}_\bfp(t)}{\partial t} 
    = \sum_{n,n'} \int d^3 q \int d^3 q'  \int_{t_0}^t d^3 x_1 \int_{t_0}^{t_1} d^4 x_2 \langle J_{(2)}^{\alpha}J_{(1)}^{\beta}\rangle \bra{i \bfp}j^{(2)}_\alpha \ket{n \bfq} \rho_\bfq^{nn'}(t) \bra{n' \bfqq} j^{(1)}_\beta \ket{j \bfp} \delta_{nn'} \delta_{ij} (2\pi)^3 \delta^{(4)}(\bfq-\bfqq) -
    \\
    - \sum_{n,n'} \int d^3 q \int d^3 q' \int_{t_0}^t d^3 x_1 \int_{t_0}^{t_1} d^4 x_2 \langle J_{(2)}^{\alpha}J_{(1)}^{\beta}\rangle  \bra{i \bfp} j_\alpha^{(1)} \ket{n \bfq} \bra{n \bfq} j_\beta^{(2)} \ket{n' \bfqq} \rho_\bfp^{n'j}(t) \delta_{in'} (2\pi)^3 \delta^{(4)}(\bfp-\bfqq) +
    \\
   +\sum_{n,n'} \int d^3 q \int d^3 q'  \int_{t_0}^t d^3 x_1 \int_{t_0}^{t_1} d^4 x_2 \langle J_{(1)}^{\alpha}J_{(2)}^{\beta}\rangle \bra{i \bfp}j^{(1)}_\alpha \ket{n \bfq} \rho_\bfq^{nn'}(t) \bra{n' \bfqq} j^{(2)}_\beta \ket{j \bfp} \delta_{nn'} \delta_{ij} (2\pi)^3 \delta^{(4)}(\bfq-\bfqq)-\\
   - \sum_{n,n'} \int d^3 q \int d^3 q' \int_{t_0}^t d^3 x_1 \int_{t_0}^{t_1} d^4 x_2 \langle J_{(1)}^{\alpha}J_{(2)}^{\beta}\rangle \rho_\bfp^{in}(t) \delta_{nj}  \bra{n \bfq} j_\alpha^{(2)} \ket{n' \bfq'} \bra{n' \bfq'} j_\beta^{(1)} \ket{j \bfp}   (2\pi)^3 \delta^{(4)}(\bfp-\bfqq)
    , 
\end{aligned}
\end{equation} 
\end{widetext}
where we account that the neutrino density matrix  is diagonal in momentum space, see (\ref{diag_p}). Here we note that the RWA is implemented automatically for those scattering processes for which $\Delta \omega = 0$.

In equation (\ref{equation_2}) there are integrals of the form:

\begin{equation}\label{int_time}
    \int d^3 \bfx_1 \int_{t_0}^{t_1} d^4 x_2 e^{i p x_1} e^{-i p x_2}.
\end{equation}
A delta function arises after integrating over the three-dimensional spatial coordinates. We consider the case where the interactions of neutrinos with the external environment are rare, allowing the use of the Markovian approximation: the lower limit of the time integral can be extended to $t_0 \rightarrow -\infty$. Then, after substituting $\tau = t_1 - t_2$, the integral (\ref{int_time}) can be written as:

\begin{equation}\label{Sohotski}
\begin{aligned}
   (2\pi)^3 \delta^{(3)}(\bfp) \int_0^\infty d & \tau e^{-i E_p \tau} =\\ &\dfrac 1 2 (2 \pi)^4 \delta^{(4)}(p) - i (2 \pi)^3 \delta^{(3)}(\bfp) P \dfrac{1}{E_p},
\end{aligned}
\end{equation}
where in the second term $P$ denotes the Cauchy principal value (see \cite{Brasil:2012trs} for details). The second term is imaginary and contributes to the loop corrections Hamiltonian $H$ (i.e., to the coherent part of neutrino evolution), which is beyond the scope of our research. 

Now consider the term $\langle J_{\nu}(x_2)J_{\mu}(x_1)\rangle$ in (\ref{equation_2}). Following \cite{Stan2025}, we can write this expression as:

\begin{equation}\label{aver}
\begin{aligned}
\langle J_{\nu}(x_2)J_{\mu}(x_1)\rangle = \sum_{\Omega} \langle \Omega |J_{\nu}(x_2)J_{\mu}(x_1)\rho_E|\Omega\rangle =\\
= \sum_{\Omega}\int\frac{d^3 k}{(2\pi)^3 2 E_{k}} \frac{1}{ 2 E_{\Omega}} \langle \Omega |J_{\nu}(x_2)| k \rangle \langle k | J_{\mu}(x_1) | \Omega \rangle \rho_E(\Omega)
\end{aligned}
\end{equation}
where $\ket{\Omega}$ is the set of states of the environment and $\rho_E$ is the density matrix of the environment. In (\ref{aver}) we use the condition
\begin{equation}
\int\frac{d^3 k}{(2\pi)^3 2 E_{k}} \frac{1}{ 2 E_{\Omega}} | k \rangle \langle k| = 1 
\end{equation}
that is valid  for not high densities of the environment. 

From (\ref{aver}) assuming that particles of environment are characterized only by their momentum $l$, we get:
\begin{equation}\label{correl}
\begin{aligned}
	\langle J_{\nu}(x_2)J_{\mu}(x_1)\rangle = \\ = \int \frac{d^3 k}{(2\pi)^3 2 E_{k}}  \frac{d^3 l}{2 E_{l}}  \bar{v}_{l} \gamma_{\nu} v_{k} e^{i x_1(l-k)} \bar{v}_{k } \gamma_{\mu}v_{l} e^{-ix_2(l -k)}  \rho_E(l).
\end{aligned}
\end{equation}

Using expressions (\ref{equation_2}), (\ref{Sohotski}) and (\ref{correl}), we obtain the final equation for neutrino evolution accounting for transitions between stationary states:
\begin{widetext}
\begin{equation}\label{Main_Evoltion}
    \begin{aligned}
        \dfrac{\partial \rho_\bfp(t)}{\partial t} 
        =  
        \left[ H (t) , \rho_\bfp(t) \right]
        - \sum_{i,j=1}^{3}\langle\sigma_{ij \:\bfp} \rangle  \left\{ \Pi_{ii}, \rho_\bfp(t) \right\} 
        + 2 \left[
        \sum_{j,i=1}^{3} \int \dfrac{d^3 \bfq}{2 (2\pi)^3 E_{\bfq j}} \dfrac{d\sigma_{j \bfq \to i \bfp}}{dq}\Pi_{ij} \rho_\bfq(t) \Pi_{ji}\right],
    \end{aligned}
\end{equation}
where $\Pi_{ij} = \ket{i}\bra{j}$ is the projector onto the neutrino state, and $\sigma_{ij \: \bfp}$ is the cross section for neutrino scattering from mass state $\ket{i}$ to mass state $\ket{j}$:

\begin{equation}\label{key}
	\begin{aligned}
		 \braket{\sigma_{ij \bfp} } =  \int \frac{d^3 k}{(2\pi)^3 2 E_{k}}  \frac{d^3 l}{ 2 E_{l}} \frac{d^3 q}{(2\pi)^3 2 E_{qi}} \: \bar{v}_{l} \gamma_{\nu} v_{k}  \bar{v}_{k } \gamma_{\mu} v_{l} 
		\cdot    \bar{u}_{p i} \Gamma^{\mu} u_{j q} \bar{u}_{q i} \Gamma^{\nu} u_{pi}  \rho_e(l) \: \delta^{4} (p + l -k -q).
	\end{aligned}
\end{equation}
We also introduce $\sigma_{j\bfq \to i \bfp}$, the cross section for neutrino scattering from state $\ket{j\bfq}$ to state $\ket{i \bfp}$:
\begin{equation}\label{key}
	\begin{aligned}
	 \braket{\dfrac{d \sigma_{j \bfq \to i \bfp}}{dq}} = \int \frac{d^3 k}{(2\pi)^3 2 E_{k}}  \frac{d^3 l}{ 2 E_{l}}   \: \bar{v}_{l} \gamma_{\mu} v_{k} \bar{v}_{k } \gamma_{\nu} v_{l} 
	\cdot  \bar{u}_{p i} \Gamma^{\mu} u_{q j} \bar{u}_{q j} \Gamma^{\nu} u_{p i} \: \rho_{e} (l)  \: \delta^{4} (p + l -k -q).
	\end{aligned}
\end{equation}
\end{widetext}

The main result of our research is the neutrino evolution  (\ref{Main_Evoltion}), which accounts for transitions between different neutrino states (the states $\ket i$ and $\ket j$ can represent mass states, flavor states, or any other neutrino states) with different momenta due to neutrino scattering on a medium. The derived equation takes the form of the Lindblad (\ref{Lindblad_eq}). We have obtained the explicit form of the dissipative operators and operators, which gives insight into its physical. This provides an opportunity to connect experimental constraints on neutrino quantum decoherence with underlying physical processes.

It is worth noting that the factors $\langle \sigma_{ij\,\mathbf{p}} \rangle$  in the obtained master equation (\ref{Main_Evoltion}) serve as decoherence parameters. Of particular interest is the fact that these decoherence parameters are proportional to the inverse neutrino mean free path in the medium. 
This allows our equation to be interpreted as describing a form of quantum 
Brownian motion of neutrinos, in which the coherent oscillation dynamics 
competes with the dissipative scattering off the environmental constituents. 
In this picture, each scattering event acts as a partial measurement of the 
neutrino flavor state, gradually destroying the off-diagonal coherences of 
the density matrix.  
\subsection{Quantum decoherence due to scattering on electrons. Neutrino quantum Zeno effect }   

This section is dedicated to a specific case of the obtained master equation (\ref{Main_Evoltion}). Specifically, we consider the effect of neutrino quantum decoherence arising from the interaction of electron neutrinos with electrons at rest in the medium. We also consider keV neutrinos which energy is less than the electron mass $E_\nu<m_e$. The neutrino interaction Hamiltonian with electrons through charged currents is \cite{Giunti:2007ry}
\begin{equation}
\begin{aligned}
H_I = \frac{G_{\mathrm{F}}}{\sqrt{2}}  \int d^3x  \overline{\nu_e} \gamma^\rho\left(1-\gamma^5\right) e \: \bar{e} \gamma_\rho\left(1-\gamma^5\right) \nu_e.
\end{aligned}    
\end{equation}
From this follows the effective neutrino interaction Hamiltonian in the flavour basis

\begin{equation}
   H_{I} = V_{CC} \: \overline{\nu}_{e} \gamma^0(1-\gamma_5) \nu_{e},
\end{equation}
where $V_{CC} = \sqrt{2} G_F N_e$. In what follows, we omit the neutrino interaction with electrons through the neutral currents since it does not contribute to the neutrino oscillations. 

Since neutrinos are ultra-relativistic particles, firstly, the neutrino energy is $E_\bfp \approx |\bfp|$ and, secondly, the neutrino scattering angle on an electron is approximately zero. Therefore, we can approximate $\rho_\bfp(t) \equiv \rho(|\bfp|, \theta_\bfp,\phi_\bfp, t) \to \rho(|\bfp|,t)$ (where $\theta_\bfp$ and $\phi_\bfp$ are neutrino propagation angles). Then the master equation (\ref{Main_Evoltion}) in the flavour basis takes the following form 
\begin{widetext}
\begin{flalign}\label{special_equation}
    \begin{aligned}
        &\dfrac{\partial \rho_\nu (|\mathbf{p}|,t)}{\partial t} = - i \left[  H(t) ,\rho_\nu (|\mathbf{p}|,t) \right]  + \left[ - \dfrac{1}{2} N_e \sigma_{ee}\left\{ \Pi_{ee}, \rho_\nu (|\mathbf{p}|,t) \right\} + N_e \int_{0}^{T_{\text{max}}} d T_e  \frac{2G_F^2m_e}{\pi} \Pi_{ee} \rho_\nu (|\mathbf{q}|,t) \Pi_{ee} \right],
    \end{aligned}&&
\end{flalign}
\end{widetext}
where $\sigma_{ee}$ is the total cross section of neutrino scattering on an electron, $\Pi_{ee} = \ket{\nu_e}\bra{\nu_e}$ , $T_e$ is the electron recoil energy and $|\mathbf{q}|$  is the momentum of neutrino after scattering that depends on the electron recoil energy. %However, it is worth noting that for keV neutrinos scattering on electrons, the maximum and minimum energies after scattering can be easily obtained. 
Since the neutrino momentum is small compared to the electron mass, it can be assumed that the neutrino momentum does not change significantly after scattering, e.g. $\rho (|\bfq|,t) \approx \rho (|\bfp|,t)$, where $\bfp$ is the neutrino momentum before scattering. Then the neutrino density matrix can be taken outside the integral in (\ref{special_equation}) and the integral becomes:
\begin{equation}\label{special_equation_integral}
    \begin{aligned}
 \int_{0}^{T_{\text{max}}} d T_e \:\frac{2G_F^2m_e}{\pi}=\frac{4 G_F^2E_{\nu}^2}{\pi}\equiv \sigma_{\text{ee}}.
      \end{aligned}
\end{equation}
Then the obtained master equation (\ref{special_equation}) reduces to the standard Lindblad equation (\ref{Lindblad_eq}) with one dissipative operator $L_1$ and corresponding parameter $\Gamma_1$:

\begin{equation}\label{dis_oper_param}
\begin{aligned}
    L_1=\Pi_{ee} \ \ \ \ \text{and} \ \ \ \ \Gamma_1 = N_e\sigma_{ee}.  \\
    \end{aligned}
\end{equation}
Thus, the neutrino master equation (\ref{special_equation}) takes the form
\begin{flalign}\label{special_equation_final}
    \begin{aligned}
        &\dfrac{\partial \rho_\nu (|\mathbf{p}|,t)}{\partial t} = - i \left[  H(t),\rho_\nu (|\mathbf{p}|,t) \right] + \\
        &\quad  +  \Gamma_1 \left(\Pi_{ee} \rho (|\mathbf{ p}|,t) \Pi_{ee} - \dfrac 1 2 \left\{ \Pi_{ee}, \rho_\nu (|\mathbf{p}|,t) \right\} \right).
    \end{aligned}&&
\end{flalign}
The Lindblad master equation can be expanded on Pauli matrices in the case of two neutrino oscillations or on Gell-Mann matrices in the case of three-flavour oscillations ($\rho(t) = \sum_n P_n F_n$ and $O= \sum_n O_n F_n$ where $F_n$ stands for Pauli or Gell-Mann matrices and $F_0$ is the unit matrix). Then the Lindblad equation can be written in the following form

\begin{equation} \label{3x3.eq.LindbladGM1}
\frac{\partial P_k (t)}{\partial t} F_k = 2\epsilon_{ijk} H_i P_j(t) F_k + D_{kl} P_l(t) F_k
,
\end{equation}
where the dissipative effects are encoded in the matrix $D_{kl}$ (indexes $k$ and $l$ starts with zero, i.e. $k,l = 0,1,2...$). For the case of two-flavour approximation in description of neutrino oscillations the dissipative matrix is given by

\begin{equation}\label{Dis_Matr_deph}
    D^{(2)}_{kl} = - \dfrac{\Gamma_1}{2}
\begin{pmatrix}
0 & 0 & 0 & 0  \\
0 & 1 & 0 & 0 \\
0 & 0 & 1 & 0 \\
0 & 0 & 0 & 0 
\end{pmatrix}
,
\end{equation}

and for the three-flavour case we get

\begin{equation}\label{dis_matrix_3}
    D^{(3)}_{kl} = - \dfrac{\Gamma_1}{2} 
\begin{pmatrix}
0 & 0 & 0 & 0 & 0 & 0 & 0 & 0 & 0 \\
0 & 1 & 0 & 0 & 0 & 0 & 0 & 0 & 0 \\
0 & 0 & 1 & 0 & 0 & 0 & 0 & 0 & 0 \\
0 & 0 & 0 & 0 & 0 & 0 & 0 & 0 & 0 \\
0 & 0 & 0 & 0 & 1 & 0 & 0 & 0 & 0 \\
0 & 0 & 0 & 0 & 0 & 0 & 0 & 0 & 0 \\
0 & 0 & 0 & 0 & 0 & 0 & 0 & 0 & 0 \\
0 & 0 & 0 & 0 & 0 & 0 & 0 & 0 & 0 \\
0 & 0 & 0 &   0 & 0 & 0 & 0 & 0 &  0
\end{pmatrix}
.
\end{equation}

The  dissipator $ D_{kl}$ in the obtained neutrino master equation (\ref{3x3.eq.LindbladGM1}) describes an important special case of the quantum decoherence: the quantum Zeno effect \cite{greenfield2025unifiedpicturequantumzeno}. 

Previously the neutrino quantum Zeno effect was considered in \cite{Daniel} only for the special case of sterile neutrinos. The obtained master equation \eqref{Main_Evoltion} is the most general equation that can describe the quantum Zeno effect between any neutrino states (e.g. mass, flavour, spin and others).

The neutrino quantum Zeno effect will occur in the limit of high electron density. In this regime, the decoherence rate $\Gamma = N_e \sigma_{ee}/2$ becomes much larger than the characteristic oscillation frequency $\Delta m^2 / 2E_\nu$, meaning that the neutrino interacts with the medium far more frequently than it oscillates. As a result, the evolution of the neutrino is no longer governed by the coherent part 
of the master equation, but is instead dominated by the Lindblad dissipator. This is precisely the condition for the quantum Zeno effect appearance: when a quantum system is measured --- or equivalently, when it interacts strongly with its environment --- faster than it can evolve, transitions between states are suppressed rather than enhanced.

In our case, the electron medium acts as a continuous measurement apparatus, effectively projecting the neutrino onto the electron flavor eigenstate at each scattering event. Consequently, 
instead of oscillating between flavors, the neutrino becomes ``frozen'' in electron state, and flavor conversion is completely suppressed. This represents a novel and striking manifestation of the quantum Zeno effect in the context of neutrino physics.

Another important result is that we have obtained the explicit form of the decoherence parameters \eqref{dis_oper_param} and corresponding  non-trivial dependence of the decoherence parameters on the neutrino energy $E_{\nu}$. This dependence can be much more complex than a simple power law and should be studied separately for different processes.

In literature most of the experimental constraints on decoherence parameters are given in the neutrino mass basis. Thus, in order to evaluate and compare the obtained results with experimental data it is necessary to switch from the flavour to the mass basis
\begin{equation}
    \rho_{m} = U^{\dagger}\rho_fU
    ,
\end{equation} 
where the mixing matrix in approximation of two-flavours is 
\begin{equation}
    U = \begin{pmatrix}
\cos \theta & \sin \theta  \\
-\sin \theta & \cos \theta
\end{pmatrix}
,
\end{equation}
where $\theta$  is mixing angle.
Then the dissipative matrix tranforms from flavour to mass basis as follows 
\begin{equation}\label{Dmkl}
   D^{m}_{kl} = \sum_{p}^{\operatorname{Dim} SU(N)}  C_{pk}D^{f}_{pl},
\end{equation}
where the matrix elements $C_{pk}$ are defined as follows 
\begin{equation}\label{eq:basis_transform}
   U^{\dagger}F_kU = \sum_{i}^{\operatorname{Dim} SU(N)}C_{ki}F_i.
\end{equation}
Then in the case of two flavors from (\ref{Dmkl}) and (\ref{eq:basis_transform}) we obtain the dissipative matrix in mass basis as follows 
\begin{equation}
    D^{m\:(2)}_{kl} = - \dfrac{N_e\sigma_{ee}}{2}
\begin{pmatrix}
0 & 0 & 0 & 0  \\
0 & \cos 2\theta  & 0 & 0 \\
0 & 0 & 1 & 0 \\
0 & \sin 2\theta & 0 & 0 
\end{pmatrix}
.
\end{equation}
Similar formulas for the transformation of the dissipative matrix between the flavor and mass bases are presented in  \cite{Dftm}. Using the values for the electron density $N_e =  5.7\cdot10^{-18} \text{ GeV}^3$  \cite{Abusleme_2025} (that corresponds to the Earth density), for the neutrino energies $E_{\nu} = 10$ keV, we obtain the decoherence parameters
\begin{equation}
\Gamma (E_{\nu})\approx \dfrac{N_e\sigma_{ee} }{2}\approx   2.5\cdot10^{-38} \text{ GeV}
\end{equation}
Taking into account the obtained expression for the decoherence parameters $\Gamma(E_\nu) = \Gamma(E_0) \left(\frac{E_{\nu}}{E_0}\right)^2$, where $E_0 = 1$ GeV, then

\begin{equation}\label{limit}
\Gamma(E_0) = \Gamma(E_\nu) \left(\frac{E_{0}}{E_\nu}\right)^2 = \dfrac{2 N_e G_F^2 E^2_{0}} {\pi} \approx 2.5 \cdot 10^{-28} \text{ GeV}
.
\end{equation}

If we compare (\ref{limit}) with experimental data, the decoherence parameter due to electron scattering is two orders of magnitude smaller than the parameters observed in the experiment \cite{Ternes:2025mys}. It should be noted, however, that a direct comparison with the reactor experiment data requires some caution. The typical energies of reactor antineutrinos are of order $\sim 1$~MeV, which lies outside the strict domain of applicability of the low-energy approximation employed in deriving equation~(\ref{special_equation}). A more careful treatment valid in this energy regime is therefore needed to be developed before 
quantitative conclusions can be drawn.

\subsection{Quantum decoherence due to scattering through NSI}

Neutrino non-standard interactions (NSI) parametrize possible deviations 
from Standard Model contributions to the effective neutrino--matter 
couplings~\cite{Farzan:2023fqa,Kling:2025zsb,Gehrlein:2025nc,Freitas:2025wsa}. Such interactions 
arise in a wide class of BSM scenarios and modify both 
neutrino propagation in matter and the contributions to effective matter couplings and detection cross 
sections. Current neutrino oscillation experiments --- including reactor 
experiments such as Daya Bay and KamLAND, and long-baseline accelerator 
experiments such as NOvA and T2K --- constrain 
NC-NSI parameters by searching for deviations from Standard Model  
predictions. 

Herein, we show that neutral-current NSI (NC-NSI) interactions with neutrons and protons give rise to neutrino quantum decoherence. In a model-independent framework, NSI are described by the following four-fermion Langrangian ~\cite{Freitas:2025wsa}:
\begin{equation}
    \mathcal{L}_{\text{NC-NSI}} = -2\sqrt{2}G_F \epsilon_{\alpha\beta}^{fX} 
    (\bar{\nu}_\alpha \gamma^\mu P_L \nu_\beta)(\bar{f} \gamma_\mu P_X f),
\end{equation}
where $G_F$ is the Fermi constant and $P_{R,L} = (1 \pm \gamma_5)/2$ 
are the chirality projection operators. The dimensionless constants  $\epsilon_{\alpha\beta}^{fX}$ quantify 
the strength of the NSI between leptons of flavours $\alpha$ and $\beta$ 
and the matter field $f \in \{e, u, d\}$ (for NC-NSI).  Notably, NC-NSI can induce 
decoherence in neutrino propagation, which is absent in the Standard 
Model neutral-current interaction.

In this section we consider neutrino evolution and oscillations in the terrestrial environment peculiar to long-baseline neutrino experiments and account for contribution of neutron and protons. For our estimations we consider NC-NSI with electrons to be zero $\epsilon^e_{ee}=0$. Then for neutrino 
energies $E_{\nu} \ll m_p$ general master equation (\ref{Main_Evoltion}) is reduced to (\ref{3x3.eq.LindbladGM1}).

\subsubsection{Lepton-flavor-conserving decoherence 
               ($\epsilon^f_{ee} \neq 0$)}

Firstly, we consider the case of neutrino scattering on neutrinos and protons when $\epsilon^f_{ee} \neq 0$. Then the dissipative matrix $D_{kl}$ takes the same form as \eqref{Dis_Matr_deph} and the decoherence parameter is
\begin{equation}
    \Gamma_{\rm NSI} = \frac{1}{2}
    \bigl( N_p \sigma^{\rm NSI}_p 
    + N_n \sigma^{\rm NSI}_n\bigr),
\end{equation}
where $N_p$ and $N_n$ are the number densities of protons and neutrons, $\sigma^{\rm NSI}_p = \frac{4 (\epsilon^{p}_{ee})^2G_F^2E_{\nu}^2}{\pi}$ and  $\sigma^{\rm NSI}_n = \frac{4 (\epsilon^{n}_{ee})^2G_F^2E_{\nu}^2}{\pi}$  are the 
corresponding neutrino scattering cross sections on  protons, neutrons. Using the ratio $Y_n = \frac{N_n}{N_p}=1.051 $ \cite{freitas2025nonstandardneutrinointeractionsneutrino}
%
%\begin{equation}\label{epsilon}
%    \epsilon_{ee} = \epsilon_{ee}^p + %Y_n\,\epsilon_{ee}^n,
%    \qquad Y_n = \frac{N_n}{N_p}=\dfrac{2}{3},
%\end{equation}
%
we get the following expression for the decoherence parameter
\begin{equation}
    \Gamma_{\rm NSI} = \frac{2}{\pi}\,((\epsilon^p_{ee})^2 + Y_n(\epsilon^n_{ee})^2 )\,G^2_F\,E_{\nu}^2\,
    N_p.
    \label{eq:GammaNSI_full}
\end{equation}

\begin{comment}

\paragraph{Neutrino scattering on electron}
Restricting to the only electron contribution and using the current 
bound $\epsilon^2_{ee} < 10^{-6}$, for $E_{\nu} = 10~\text{keV}$ and 
$N_e = 5.7\cdot10^{-18}~\text{GeV}^3$~\cite{Abusleme_2025}. From (\ref{eq:GammaNSI_full}), the decoherence parameter 
is bounded by

%
\begin{equation}
    \Gamma^{e}_{\rm NSI}(E_0) = \frac{2}{\pi}\,\epsilon^2_{ee}\,G^2_F\,
    N_e\,E^2_0 \cos(2\theta) < 2\cdot10^{-34}~\text{GeV}
\end{equation}
%
This value is many orders of magnitude below current experimental 
sensitivity, indicating that decoherence via NSI electron scattering 
is negligible at keV energies.
\end{comment}

Using experimental constraints on decoherence parameters obtained from long-baseline experiments $\Gamma_{\text{exp}}(E_0) < 1.3 \times 10^{-26} $ GeV \cite{DeRomeri:2023dht}, where $E_0 = 1$ GeV,
together with the nucleon density 
$N_p = 5.7 \times 10^{-18}~\text{GeV}^3$~\cite{Abusleme_2025}, 
the following estimation on NC-NSI constants are obtained:
\begin{equation}
  (\epsilon^p_{ee})^2 + Y_n(\epsilon^n_{ee})^2 < \frac{\pi}{2\,G^2_F\,E_0^2\,N_p} 
    \,\Gamma_{\rm exp}(E_0) \approx 20,
\end{equation}
from which it follows that
\begin{equation}\label{epsilonp}
    -4.47 < \epsilon^p_{ee} < 4.47, 
\end{equation} and
\begin{equation}\label{epsilonn}
    -4.36 < \epsilon^n_{ee} < 4.36.
\end{equation}
\begin{comment}
It is useful to use the coefficient $\epsilon_{ee} = \epsilon_{ee}^p + Y_n\,\epsilon_{ee}^n$ in order to compare the obtained estimations with existing constraints. The our estimation is
\begin{equation}
   -5.77< \epsilon_{ee} < 5.77
   .
\end{equation}
\end{comment}
The obtained estimations agree with dedicated experimental 
constraints~\cite{Gehrlein:2025nc}.

\subsubsection{Lepton-flavor-violating decoherence 
               ($\epsilon^f_{e\mu} \neq 0$)}

The influence of lepton-flavor-violating NC-NSI on neutrino quantum 
decoherence is now considered. For $\epsilon_{e\mu} \neq 0$, the corresponding dissipator matrix in the flavour basis takes the form
\begin{equation}
    D^{f}_{kl} = -\Gamma_{\text{NSI}}
    \begin{pmatrix}
    0 & 0 & 0 & 0 \\ 0 & 1 & 0 & 0 \\ 
    0 & 0 & 1 & 0 \\ 0 & 0 & 0 & 2
    \end{pmatrix}.
\end{equation} Here
$\Gamma_{\text{NSI}}= \frac{N_n\sigma^{n}_{e\mu} + N_p\sigma^{p}_{e\mu}}{2}$,
where $\sigma^{p}_{e\mu} = \frac{4G_F^2 E_\nu^2}{\pi} (\epsilon^{p}_{e\mu})^2$ and $\sigma^{n}_{e\mu} = \frac{4G_F^2 E_\nu^2}{\pi} (\epsilon^{n}_{e\mu})^2$ denote the NC-NSI-induced cross section for neutrino scattering on protons ($\nu_e + p \to \nu_{\mu} + p$) and neutrons ($\nu_e + n \to \nu_{\mu} + n$) correspondingly.
After switching  $ D^{f}_{kl}$ to the mass basis using ~(\ref{eq:basis_transform}) we get
\begin{equation}
    D^{m}_{kl} = - \Gamma_{\text{NSI}}
    \begin{pmatrix}
    0 & 0 & 0 & 0 \\ 
    0 & \cos2\theta & 0 & \sin2\theta \\ 
    0 & 0 & 1 & 0 \\ 
    0 & -\sin2\theta & 0 & 2\cos2\theta
    \end{pmatrix}.
\end{equation}
In the same way as for the previous case, the following estimations are obtained 
%
\begin{comment}
\begin{equation}
    (\epsilon^p_{e\mu})^2 + Y_n (\epsilon^n_{e\mu})^2 < \frac{\pi}{2\,G^2_F\,E_0^2\,N_p} 
    \,\Gamma_{\rm exp}(E_0) \approx 20,
\end{equation}
\end{comment}
%
from which it follows that
\begin{equation}\label{epsilonp}
    -4.47 < \epsilon^p_{e\mu} < 4.47, 
\end{equation} and
\begin{equation}\label{epsilonn}
    -4.36 < \epsilon^n_{e\mu} < 4.36.
\end{equation}

These results and estimations demonstrate that neutrino quantum decoherence can serve as a pathway to constrain neutrino non-standard interactions, including those responsible for lepton flavor violation. \cite{Farzan:2023fqa,Kling:2025zsb,Gehrlein:2025nc,Freitas:2025wsa}.

\subsection{Quantum decoherence due scattering on dark matter fermions}

The obtained master equation (\ref{special_equation}) can also be employed to describe neutrino decoherence induced by interactions with fermionic dark matter \cite{universe9040197,dev2025newconstraintsneutrinodarkmatter,PhysRevD.97.075039}. The interaction Lagrangian density accounting for the coupling of electron neutrinos with fermionic dark matter mediated by the vector $Z'$-boson is
\begin{equation}
   \mathcal{L}_{\mathrm{int}} = -g_{\chi_{\mathrm{L}}} \overline{\chi_{\mathrm{L}}} \gamma^\mu Z_\mu^{\prime} \chi_{\mathrm{L}} - g_\nu \overline{\nu_e} \gamma^\mu Z_\mu^{\prime} \nu_{e},
\end{equation} where $g_{\chi_{\mathrm{L}}}$ is the coupling constant of the dark matter field with the $Z'$-boson, $g_\nu$ is the coupling constant of the electron neutrino with the $Z'$-boson,  $\chi_{\mathrm{L}}$ is the fermionic dark matter field. Then the effective  Hamiltonian that describes neutrino scattering with dark fermions reads
\begin{equation}
   H_{\mathrm{int}} = \int d^3 x \dfrac{g_{\chi_{\mathrm{L}}}g_\nu  }{m^2_{Z^{\prime}}} \overline{\chi_{\mathrm{L}}} \gamma^\mu \chi_{\mathrm{L}} \overline{\nu_e} \gamma_\mu  \nu_{e}
   .
\end{equation} 

For fermionic dark matter with a mass of about $m_{DM} = 100$ MeV and neutrino energy $E_{\nu} < m_{DM},$ we can use the master equation (\ref{3x3.eq.LindbladGM1}) and  dissipative matrix  \eqref{Dis_Matr_deph}. Then for neutrino decoherence parameter due to scattering on dark matter fermions we get
\begin{equation}
       \Gamma_{DM}(E_\nu) = N_{DM} \frac{g_\chi^2 g_{\nu_{e}}^2}{4 \pi} \frac{ E_\nu^2}{m_{\mathrm{Z}^{\prime}}^4}
       ,
\end{equation}
where $N_{\rm DM}$ is the dark matter number density, $m_{Z'}$ is 
the mass of the $Z'$-boson.

Following discussion from \cite{dev2025newconstraintsneutrinodarkmatter} we take $m_{Z^{\prime}} =3\:m_{DM} = 300 $ MeV,  $g_\chi g_{\nu_{e}} < 10^{-4} $, and accounting current constrains   on $N_{DM} < 0.5 \:\dfrac{\text{GeV}}{\text{cm}^3}$ \cite{PhysRevD.104.083023} , and $E_{\nu} = 1$ MeV we get  \begin{equation}
       \Gamma_{DM}(E_0)  <  10^{-44} \text{ GeV}.
\end{equation}
This estimation demonstrates that dark matter scattering contributes negligibly to neutrino quantum decoherence, falling many orders of magnitude below current experimental sensitivity.

\section*{CONCLUSION}

A quantum field-theoretic framework for describing neutrino evolution in a fermionic medium has been developed. We have shown that neutrino
quantum decoherence can be caused by neutrino scattering. This effect has been considered for the first time. The derived master 
equation~(21), which accounts for momentum-changing transitions 
between different neutrino states, represents a significant 
generalization of the Lindblad master equation and explicitly 
connects the decoherence parameters to scattering cross 
sections. 

The developed framework has been applied to several physically 
motivated scenarios: 1) neutrino scattering on electrons, 2) neutrino scattering on protons and neutrons through NC-NSI, 3) neutrino scattering on fermionic dark matter. In the first scenario a novel manifestation of the quantum Zeno effect in neutrino physics has been identified. In the second scenario, estimations on the NC-NSI constants have been derived directly from quantum decoherence measurements. This demonstrates that decoherence experiments can serve as a complementary probe of physics beyond the Standard Model. In the third scenario we have  showed that contribution of fermionic dark matter scattering to decoherence is negligible, falling many orders of magnitude below current experimental sensitivity.

It is worth noting that scattering processes exhibit high sensitivity to potential new physics beyond the Standard Model. As a result, numerous current and future experiments—such as COHERENT \cite{Akimov2017}, SATURNE \cite{Kouzakov:2025duy, Cadeddu:2019qmv}, HeRald \cite{SPICE:2023tru} and other are dedicated to investigating neutrino scattering phenomena. The importance of the presented results lies in demonstrating that neutrino scattering processes can also be studied in neutrino oscillation experiments through the neutrino quantum decoherence.

\section*{Acknowledgement}
The authors are thankful to Konstantin Kouzakov and Fedor Lazarev  for helpful comments on the manuscriptfor and valuable discussions.
This work was supported by the Russian Science
Foundation (project no. 24-12-00084).

\bibliography{ver3}

\end{document}